\documentclass[pre,aps,twocolumn]{revtex4}
\usepackage{amsmath}
\usepackage{graphicx}
\begin{document}

\title{Rich Variety of Bifurcations and Chaos in a Variant of
 Murali-Lakshmanan-Chua Circuit}

\author{K.Thamilmaran} 

\affiliation{Centre for Nonlinear Dynamics, Department of Physics,
Bharathidasan University, Tiruchirapalli 620024, India}

\author{K. Murali}  

\affiliation{Department of Physics, Anna University, Chennai 
600025, India} 

\author{M. Lakshmanan}  
\email{lakshman@bdu.ernet.in}
\affiliation{Centre for Nonlinear Dynamics, Department of Physics,
Bharathidasan University, Tiruchirapalli 620024, India}

\date{ }

\begin{abstract}

A very simple nonlinear \emph{parallel} nonautonomous LCR circuit with
Chua's diode as its only nonlinear element, exhibiting a rich variety
of dynamical features, is proposed as a variant of the simplest
nonlinear nonautonomous circuit introduced by Murali, Lakshmanan and
Chua(MLC).  By constructing a two-parameter phase diagram in the
$(F-\omega)$ plane, corresponding to the forcing amplitude $(F)$ and
frequency $(\omega)$, we identify, besides the familiar period-doubling
scenario to chaos,  intermittent and quasiperiodic routes to chaos as
well as period-adding sequences, Farey sequences, and so on.  The
chaotic dynamics is verified by both experimental as well as computer
simulation studies including PSPICE.

\end{abstract}

\maketitle
\section{Introduction}

Sometime ago, the simplest nonlinear dissipative nonautonomous
electronic circuit consisting of a forced  {\it{\underline{series}}}
LCR circuit connected parallely to the Chua's diode, which is a
nonlinear resistor, was introduced by Murali, Lakshmanan and
Chua~[Murali et al., 1994].  The circuit exhibits several interesting
dynamical phenomena including period doubling bifurcations, chaos and
periodic windows.  However, in the paramater regimes investigated, it
does not exhibit other important features such as quasiperiodicity,
intermittency, period adding sequences, and so on.  It will be quite
valuable from nonlinear dynamics point of view to construct a simple
electronic circuit which exhibits a wide spectrum of dynamical
phenomena~[Lakshmanan~\&~Murali, 1996].  In this paper, we wish to
point out that a rich variety of phenomena can be realized with a
simple variant of the above MLC circut, by connecting the Chua's diode
to a forced {\it{\underline{parallel}}} LCR circuit instead of the
forced {\it{\underline{series}}} LCR circuit.  The resultant system
exhibits not only the familar period doubling route to chaos and
windows but also intermittent and quasiperiodic routes to chaos as well
as period adding sequences and Farey sequences, to name a few.

\section{CIRCUIT REALIZATION OF MLC VARIANT CIRCUIT}

\begin{figure}[ht!]
\begin{center}
\includegraphics[width=0.98\linewidth]{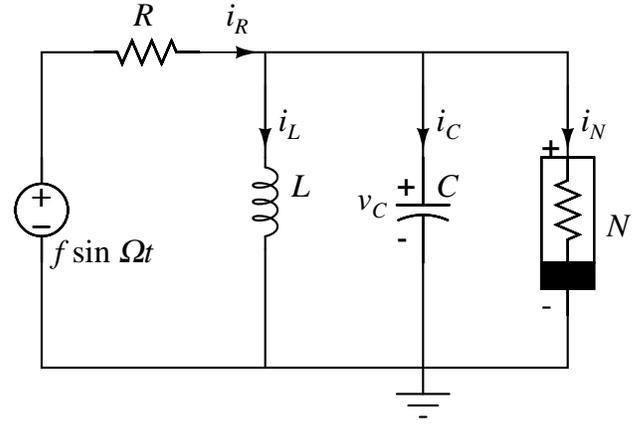}
\end{center}
\caption{Circuit realization of the simple non-autonomous MLC variant
circuit.  Here $N$ is the Chua's diode.}
\label{fig1}
\end{figure}

The circuit realization of the proposed simple non-autonomous circuit,
namely, a variant of the standard MLC  circuit, is shown in
Fig.~\ref{fig1}. It consists of a capacitor($C$), an inductor($L$), a
resistor($R$), an external periodic forcing voltage source and only one
nonlinear element($N$), namley, the Chua's diode. In the dynamically
interesting range, the $v-i$ characteristic of the Chua's diode is
given by the usual three segment piecewise-linear function~[Chua et
al., 1987; Kennedy,~M.P, 1992; Cruz, J.M \& Chua, L.O, 1992]. The
nonlinear element is added to the familiar forced
{\it{\underline{parallel}}} LCR circuit instead of the
\underline{series} LCR of MLC circuit [Murali et al., 1994]. The
resulting circuit can be considered as another important very simple
dissipative second order nonautonomous nonlinear circuit and a variant
of the MLC circuit.  By applying Kirchhoff's laws to this circuit, the
governing equations for the voltage $v$ across the capacitor $C$ and
the current $i_L$ through the inductor $L$ are represented by the
following set of two first-order non-autonomous diffrential equations:
\begin{subequations}
\label{eq:1}
\begin{eqnarray} C \frac{dv}{dt}  & = & \frac{1}{R} (f \sin( \Omega
t)-v) -i_L-g(v), \label{eq:1a}\\ 
L \frac{di_L}{dt} & = & v, \label{eq:1b} 
\end{eqnarray}
\end{subequations} 
where $g(.)$ is a piecewise-linear function defined by Chua et al.  and
is given by $g(v)=G_bv+0.5(G_a-G_b)[|v+B_p|-|v-B_p|]$ which is the
mathematical representaion of the charactierstic curve of Chua's diode
\begin{figure}[ht!]
\begin{center}
\includegraphics[width=\linewidth]{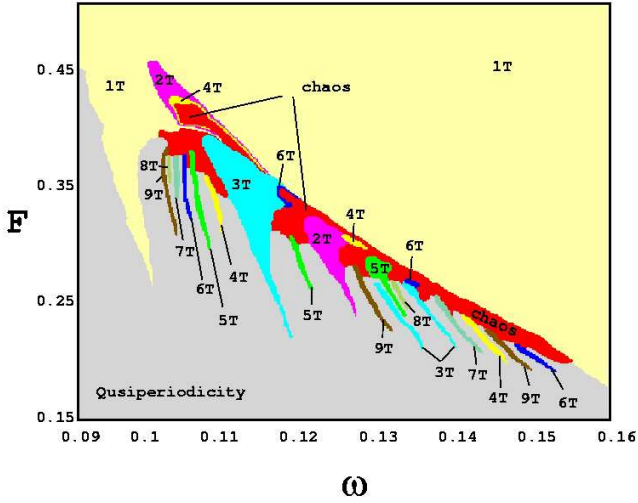}
\end{center}
\caption{Two parameter phase diagram in the ($F-\omega$) plane.}
\label{fig2}
\end{figure}
[Kennedy,~M.P, 1992; Cruz, J.M \& Chua, L.O, 1992].  The values of the
various circuit parameters chosen for our study are as follows:
$C$={10.15 nF}, $L$={445 mH}, $R$={1745 ohms}, $\Omega =1.116 KHz$,
$G_a=-0.76ms$,  $G_b=-0.41ms$, and $B_p=1.0 V$.

In equation(1),  $f$ is the amplitude and $\Omega$ is the angular
frequency of the externl periodic signal.  Rescaling Eq.~(\ref{eq:1})
as $v=xB_p$, $i_L=GyB_p$, $G=1/R$, $\omega =\Omega C/G$, and $t =
\tau C/G$ and then redefining $\tau$  as $t$, the following set of
normalized equations are obtained:
\begin{subequations}
\label{eq:2}
\begin{eqnarray}
\dot{x} & = & F \sin(\omega t)-x-y-g(x),
\label{eq:2a}\\ 
\dot{y} & = & \beta x,\qquad\qquad \left( .= \frac{d}{dt} \right) 
\label{eq:2b}
\end{eqnarray}
\end{subequations} 
where $\beta = C/L G^2$, $F=f /B_p$, $G=1/R$.  Obviously
$g(x)=bx+0.5(a-b)(|x+1| - |x-1|)$.  Here $a =G _a/G$, $b=G_b/G$.  Note
that the two coupled first order ordinary differential equation given
by Eqs.~(\ref{eq:2}) can also be written as a single second order
differential equation of the Lienard's type in the form
\begin{equation} 
\ddot{y}+\dot{y}+\beta g(\dot{y}/\beta)+\beta {y}  
= \beta F \sin(\omega t). 
\label{eq:3}
\end{equation}
We note at this point chaos via torus breakdown generated in a
piecewise-linear forced van der Pol equation of the form(3) with
asymmetric nonlinearity has been studied by Inaba and Mori [Inaba, \& Mori, 1991] 
sometime ago.  However the corresponding circuit uses more nonlinear
elements than the present circuit.  Now the dynamics of equation$(2)$
or equivalently $(3)$ depends on the parameters $a$, $b$, $\beta$,
$\omega$ and $F$.  Then for the above chosen experimental circuit
parameter values, we have $\beta=0.05$, $a=-1.121$, $b= -0.604$ and
$\omega = 0.105$.  We use the amplitude $F$ of the external periodic
forcing as the control parameter and carry out a numerical simulation
of Eqs.~(\ref{eq:2}) either by integrating  Eqs.~(\ref{eq:2}) or by
solving Eq.~(\ref{eq:3}) analytically and numerically, for increasing
values of $F$.  For the above choice of parameters the numerical
simulation of Eqs.~(\ref{eq:2}) exhibits novel dynamical phenomena.
\begin{figure}[ht!]
\begin{center}
\includegraphics[width=\linewidth]{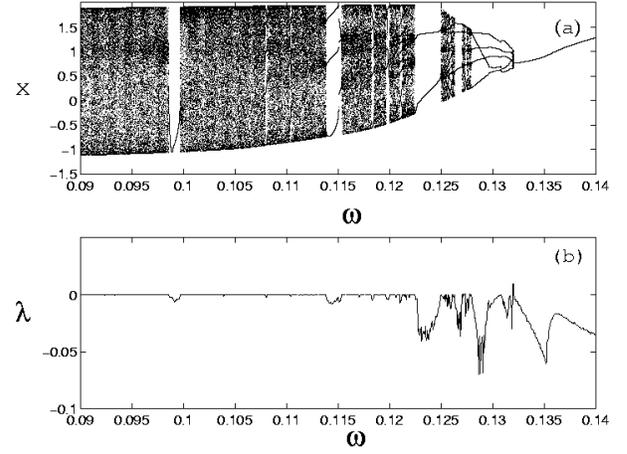}
\end{center}
\caption{(a) One parameter bifurcation diagram in the ($\omega-x$)
plane and (b) maximal Lyapunov spectrum ($\lambda_{max}$).  The value
of the forcing amplitude has been fixed at $F=0.28$ }
\label{fig3}
\end{figure}

To begin with we have carried out an experimental study of the dynamics
of the circuit given by Fig.~\ref{fig1}. The driving amplitude($f$) is
slowly increased from 0V and the response of the system is observed to
progress through a series of transition from periodic motion to
aperiodic motion.  When the driving amplitude $f=0$ (corresponding to
the autonomous case), a period-1 limit cycle is observed.  By
increasing the amplitude from zero upwards, the circuit of
Fig.~\ref{fig1} is found to exhibit a quasiperiodic(torus) attractor
which then transits to chaos via torus breakdown, followed by periodic
windows, period-adding sequences etc.  We have confirmed these results
also by using the standard 4th-order Rugne-Kutta integration routine
and carrying  out a numerical analysis of Eqs.~(\ref{eq:2})  or
equivalently Eq.~(\ref{eq:3}) with the rescaled circuit parameters of
Fig.~\ref{fig1} as given above.  We have observed a series of
appearence and disappearence of the quasiperiodic (torus) and phase
locking(periodic) attractors, and period-adding sequences alternatively
by varying the amplitude $F$ of the external source at fixed
frequency.  Illustrated in Fig.~\ref{fig2} is the two paramater phase
diagram which is plotted in the ($F-\omega$) plane.   In particular,
quasiperiodic motions and period-adding sequences besides the standard
bifurcation sequences have been observed in regions of the lower drive
amplitude ($F$) and frequency ($\omega$) values. For example, typical
quasiperiodic attractors exist in the range  $F= (0.15,0.2)$, and
$\omega$ =$(0.09,0.16)$.   Similarly, a period adding sequence exists
for $F= (0.36,0.38)$ and $\omega=(0.1,0.115)$, a period-doubling
bifurcation sequence for $F= 0.39$, and $\omega= (0.09,0.1)$,  and also
 type I intermittency has been identified for $F=0.38$ and  $\omega=
(0.10636)$ . In addition for $F= (0.24,0.27)$ and  $\omega =
(0.138,0.146)$, as well as for $F= (0.26,0.29)$, $\omega = (0.128,
0.138)$ and for $F= (0.28,0.33)$, $\omega  =(0.11, 0.125)$,  Farey
sequences [Kaneko, 1986] exist.  The regions of chaos are also
indicated in Fig.~\ref{fig2}.  Finally Fig.~\ref{fig3}(a).  represents
the one-parameter bifurcation in the $(\omega-x)$ plane, for $F=0.28$
which consists of quasiperiodicity, chaos, windows, period adding
sequences and the familiar period doubling bifurcation sequence,
intermittency and so on.   In fig.~\ref{fig3}(b) its corresponding
maximal Lyaponov spectrum in $(\omega-\lambda_{max})$ is plotted. Also
\begin{figure}
\begin{center}
\includegraphics[width=\linewidth]{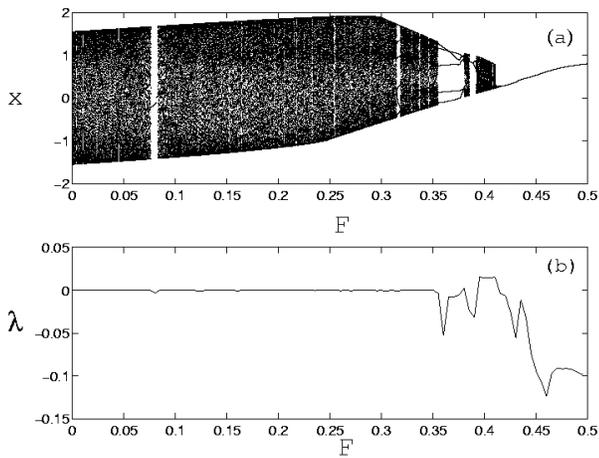}
\end{center}
\caption{(a) One parameter bifurcation diagram in the ($F-x$) plane and
(b) maximal Lyapunov spectrum ($\lambda_{max}$).  The value of the
forcing frequency has been fixed at $ \omega=0.105$}
\label{fig4}
\end{figure}
fig.~\ref{fig4}(a) represents the one-parameter bifurcation in the
$(F-x)$ plane, for $\omega=0.105$, fig.~\ref{fig4}(b) depicts the
corresponding maximal Lyaponov spectrum $(F-\lambda_{max})$ .  Further,
the experimental results obtained for a choice of circuit parameters
were also confirmed by using the PSPICE [Roberts, 1997] deck available
to simulate the behavior of the circuit in Fig.~\ref{fig1}.  In
Fig.~\ref{fig5}, we have included for comparison the chaotic attractor
corresponding to $F=0.389V$ in Eq.~(\ref{eq:1}) for experimental and
SPICE analysis and corresponding numerical simulation for amplitude
$f=0.383$ in Eq.~(\ref{eq:2}).
\begin{figure}[ht!]
\begin{center}
\includegraphics[width=0.6\linewidth]{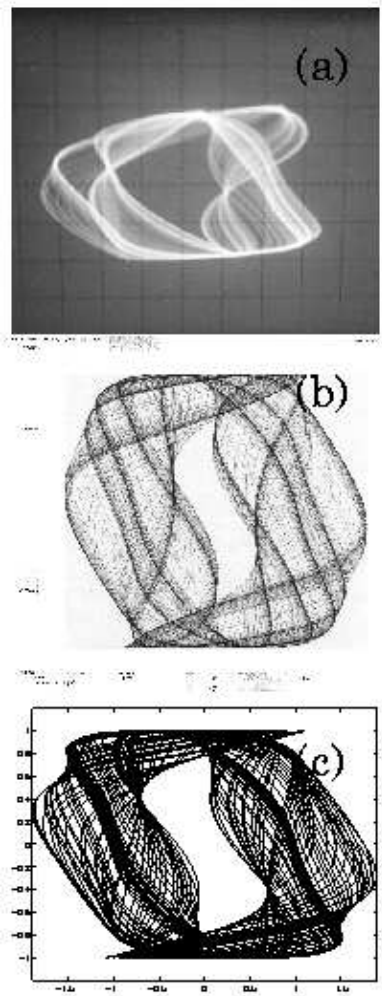}
\end{center}
\caption{(a) Chaotic attractor corresponding to $F=0.389V$, $\Omega
=1.116KHz$ in Eq.~(\ref{eq:1}): (a) experimental (b) PSPICE  and (c)
corresponding numerical simulation of Eq.~(\ref{eq:2})}.
\label{fig5}
\end{figure}

\section {CONCLUSION} 

In this letter, we have presented a very simple second order dissipative
nonlinear nonautonomous circuit which is a variant of the MLC circuit and
carried out.  It is shown to possess a very rich variety of dynamical phenomena. 
In view of the appearance of several ubiquitous routes to chaos such as
quasiperiodicity, intermittency, period doubling, period-adding and
Farey sequence in such a single but simple circuit, one can make use of the
circuit in  varied investigations on chaotic dynamics and
applications, including spatio-temporal studies.  A more detailed
analysis of these aspects will be published elsewhere.

\begin{acknowledgments}
The work of M.Lakshmanan forms part of a Department of Science and
Technology, Government of India research project. We are thankful to
Mr. A. Venkatesan for his assistance for numerical analysis.
\end{acknowledgments}

\section*{REFERENCES} 
\begin{description}
\item
Chua,~L.~O., ~Desoer,~C.A. ~\& ~Kuh,E.~S. ~$[1987]$ ~{\it{Linear and Nonlinear
Circuits}} ~(McGraw-hill, New York).

\item
Cruz,~J.~M, ~\& ~Chua, L.~O. ~$[1992]$ ~``A CMOS IC nonlinear reisitor for Chua's
circuit," {\it{IEEE Trans. Circuits Syst-I}}., vol.39, pp.985-995.

\item 
Inaba,~N.  ~\& ~Mori,~S. ~$[1991]$ ~ ``Chaos via torus break down in a piecwise-linear
forced van der Pol oscillator with a diode," {\it{IEEE Trans. Circuits
and Syst-I}}., vol.38, pp.398-409.

\item
Kaneko,~K. ~$[1986]$ ~{\it Collapse of Tori and Genesis of Chaos in Dissipative
Systems} (World Scientific,  Singapore).

\item
Kennedy,~M.~P. ~$[1992]$ ~``Robust op amp realization of Chua's circuit,"
{\it{Frequanz}}, vol.46, pp.66-80.

\item  
Lakshmanan,~M. ~\& ~Murali~K. ~$[1996]$ ~{\it{Chaos in Nonlinear Oscillators:
Controlling and Synchronization}}  (World Scientific, Singapore) .

\item 
Murali,~K., ~Lakshmanan, M. ~\& ~Chua,~L.~O. ~$[1994]$ ~``The simplest dissipative
nonautonomous chaotic circuit," {\it{IEEE Trans.  Circuits Syst-I}}.,
vol.41, pp.462-463.

\item 
Murali,~K., ~Lakshmanan, M. ~\& ~Chua,L.~O. ~$[1994]$ ~``Bifurcation and chaos in the
simplest dissipative nonautonomous circuit," {\it{Int.J.Bifurcation and
Chaos}}, vol.4, pp.1511-1524. 

\item
Roberts,~W. ~$[1997]$ ~{\it{SPICE}} (Oxford University Press, New York).
\end{description}

\end{document}